\documentclass[10pt,a4paper,onecolumn]{article}
\usepackage[a4paper, total={6in, 8in}]{geometry}
\usepackage{setspace,url}
\usepackage{lineno,hyperref,multirow}

\usepackage{epsfig,amssymb,amsmath}
\usepackage[utf8]{inputenc}
\usepackage{amsmath,graphicx}
\usepackage{amssymb,amsmath,epsfig,url,mathrsfs} \usepackage{algorithm,algorithmic}
\usepackage[usenames, dvipsnames]{color}
\usepackage[framemethod=TikZ]{mdframed}
\usepackage{xcolor}
\usepackage{colortbl}
\usepackage[colorinlistoftodos]{todonotes}
\usepackage{enumitem}
\usepackage{verbatim}
\usepackage{titling}
\usepackage{xcolor}
\usepackage{listings}
\usepackage[raggedrightboxes]{ragged2e}
\usepackage{booktabs}
\usepackage{graphicx}
\usepackage{subfig}

\usepackage[style=ieee]{biblatex}
\usepackage{amsthm}
\usepackage{etoolbox}

\makeatletter
\newcommand{\mathleft}{\@fleqntrue\@mathmargin0pt}
\newcommand{\mathcenter}{\@fleqnfalse}
\makeatother
 
\theoremstyle{definition}

\hypersetup{
   unicode=false,          
   pdftoolbar=true,        
   pdfmenubar=true,        
   pdffitwindow=false,     
   pdfstartview={FitH},    
   pdftitle={WildSpoof Evaluation Plan},    
   pdfauthor={WildSpoof},     
   pdfsubject={WildSpoof},   
   pdfcreator={},   
   pdfproducer={}, 
   pdfkeywords={automatic speaker verification and countermeasures challenge, spoofing, presentation attack detection}, 
   pdfnewwindow=true,      
   colorlinks=true,       
   linkcolor=blue,          
   citecolor=blue,        
   filecolor=magenta,      
   urlcolor=blue           
}

\mdfdefinestyle{MyFrame}{%
    linecolor=black,
    outerlinewidth=.1pt,
    roundcorner=1pt,
    innertopmargin=\baselineskip,
    innerbottommargin=\baselineskip,
    innerrightmargin=10pt,
    innerleftmargin=-10pt,
    backgroundcolor=green!10!white} 

\newtoggle{showPhaseZero}
\newtoggle{showPhaseOne}
\newtoggle{showPhaseTwo}

\togglefalse{showPhaseZero}

\togglefalse{showPhaseOne}

\toggletrue{showPhaseTwo}

\title{\textbf{WildSpoof Challenge Evaluation Plan}\thanks{\textcolor{blue}{Document Version 0.1 (\today)}}}

\author{
Yihan Wu,
Jee-weon Jung,
Hye-jin Shim,
Xin Cheng,
Xin Wang\\ \url{https://wildspoof.github.io}
}

\begin{document}

\maketitle


\section{Introduction}
\label{sec:introduction}

The WildSpoof Challenge aims to advance the use of in-the-wild data in two intertwined speech processing tasks.
It consists of two parallel tracks: 
\begin{enumerate}
\item Text-to-Speech (TTS) synthesis - generating spoofed speech, and 
\item Spoofing-robust Automatic Speaker Verification (SASV) - detecting spoofed speech.
\end{enumerate}
The organizers coordinate both tracks and define the data protocols, while participants treat the tracks as separate and independent.

The primary objectives of the challenge are:

\begin{itemize}
\item Promote in-the-wild data usage for both TTS and SASV, moving beyond conventional clean and controlled datasets, considering real-world scenario
\item Encourage interdisciplinary collaboration between spoofing generation (TTS) and detection (SASV) communities, fostering the development of more integrated, robust, and realistic systems.
\end{itemize}


\section{Challenge guideline -- TTS track}
\label{sec:guidelines}

The TTS track requires participants to build a TTS system that generates speech in the voices of specific target speakers. Given input text and one or a few utterances from a target speaker, the TTS system should generate a speech that accurately reads the text in the target speaker's voice.

\subsection{Training data}

The TTS track adopts the training set of the TITW dataset~\cite{jung2025text}. TITW has two training sets, TITW-Easy and TITW-Hard. Participants are allowed to utilize either of them or both. The file lists are listed in Table~\ref{tab:protocol_list}.



\subsection{Evaluation data}

The TTS track evaluation data is listed in the evaluation protocols of the TITW dataset: 
\begin{itemize}
\item TITW-KSKT (Known Speaker, Known Text) - \texttt{tts\_test\_KSKT.csv} 
\item TITW-KSUT (Known Speaker, Unknown Text) - \texttt{tts\_test\_KSUT.csv}. 
\end{itemize}

Both evaluation protocols cover 40. While TITW-KSKT uses text present in the training data, TITW-KSUT uses unseen text.

\subsection{What and how to submit}

Participants should submit a single zip package that includes 9,113 speech audio files for TITW-KSKT and 8,000 audio files for TITW-KSUT, respectively. Note that
\begin{itemize}
    \item File name must follow the one specified in the TTS evaluation protocol.
    \item All speech files should be in \textbf{.wav} format.
    \item All files should have a \textbf{16~kHz} sample rate with \textbf{16 bits} per sample and \textbf{PCM} encoding. The encoding can be verified using the \texttt{soxi} or equivalent utility: 
    \begin{lstlisting}[language=bash]
$: soxi file.wav
Input File     : `file.wav'
Channels       : 1
Sample Rate    : 16000
Precision      : 16-bit
Duration       : 00:00:05.32 = 85121 samples
File Size      : 84.3k
Bit Rate       : 127k
Sample Encoding: 16-bit Signed Integer PCM
    \end{lstlisting}    
    \item The archive should be compiled without a parent directory structure. You may use command below to check:
    \begin{lstlisting}[language=bash]
$: zip -r submission.zip -j directory_of_wav
$: unzip -l submission.zip
file1.wav
file2.wav
...
    \end{lstlisting}
\end{itemize}

\subsection{Metrics}

MCD, UTMOS, DNSMOS, WER, SPK-sim. All these metrics will be calculated by \href{https://github.com/wavlab-speech/versa}{Versa}.

There will be no ranking of the TTS systems.

\begin{table*}[t!]
    \centering
    \caption{Protocols and metadata files. ``File list name'' corresponds to the names of files of the downloaded TITW and SpoofCeleb datasets.}
    {
    \begin{tabular}{rll}
    \toprule
    Track & Partition & File list name \\
    \midrule
    \multirow{3}{*}{TTS}  & \multirow{2}{*}{Training} & \texttt{bonafide\_metadata\_cfg\_v3/wav.scp} and/or\\ 
    & & \texttt{bonafide\_metadata\_cfg\_v6/wav.scp}\\
    & Development & Left up to participants to use part of the training partition\\
    & Evaluation & to be released\\
    \midrule
    \multirow{3}{*}{SASV}  
    & Training & \texttt{metadata/train.csv} \\
    & Development & \texttt{protocol/sasv\_development\_evaluation\_protocol.csv} \\
    & Evaluation & to be released  \\
    \bottomrule
    \end{tabular}
    }
    \label{tab:protocol_list}
\end{table*}

\subsection{Rules}
\begin{itemize}
    \item 
    Participant who registers for the TTS track CANNOT participate in the SASV track.
    \item 
    Participants CAN use pre-trained codecs or models, but they have to be fine-tune with TITW at the last phase (if they are trainable).
    \item No ranking will be provided for the TTS track. However, 
    a summary of the challenge results will also be prepared by the organizers. Any participant who wish to retain anonymity should provide an anonymous team identifier with their registration. 
    \item 
    Participants commit to respecting the guidelines detailed earlier in this document.  Track, condition, and data/resource use guidelines should also be interpreted as challenge rules.  If a participant is uncertain as to whether any particular data or resource is permitted, they are advised to seek clarification from the organizers.
\end{itemize}
The organizers reserve the right to exclude systems from ranking and to exclude teams from future participation in WildSpoof in the event that the above rules are not adhered to.

\section{Challenge guideline -- SASV track}

The SASV track involves building an SASV system, which must compare an unlabeled probe (test) utterance to an enrollment utterance(s) of known target speakers. 
The system will be benchmarked using a mix of three types of trials---\emph{target} (bonafide target), \emph{non-target} (bonafide non-target), and \emph{spoof} (spoofed target). The SASV system should accept target trials only, rejecting the utterances that are either spoofed or do not match the target speakers' voices. 

\subsection{Training data}

The training data of the SASV track is the same as the SpoofCeleb training data~\cite{jung2025spoofceleb}. The file lists are listed in Table~\ref{tab:protocol_list}.

\subsection{Evaluation data}

The evaluation data will consist of a package of bona fide and spoofed speech waveform files. An evaluation trial list will be shared, consisting of three types of trials---\emph{target} (bonafide target), \emph{non-target} (bonafide non-target), and \emph{spoof} (spoofed target).

Each row of the trial list specifies the test utterance and the enrolment target speaker ID. 
\begin{lstlisting}[language=bash]
    enroll test 
    spk1   utt1  
    spk1   utt2  
    spk1   utt3  
    ...
\end{lstlisting}

\subsection{What to submit}

A \textbf{tsv} file should be submitted, each row of which specifies the trial name, the enrolment target speaker ID, and the score produced by the SASV system. 
\begin{lstlisting}[language=bash]
    enroll  test score
    spk1    utt1  3.66776
    spk1    utt2  1.26782
    spk1    utt3  3.29123
    ...
\end{lstlisting}  

For each row, the SASV system should produce a continuous-valued score indicating how likely the trial is \emph{target}. A larger score indicates that the trial is more likely to be \emph{target}. There is no need to constrain the score to be within 0 and 1. 

\subsection{Metric}
The evaluation metric is the \emph{agnostic} DCF (a-DCF)~\cite{shim2024adcf} which assigns a cost to a system, taking into account the miss rate and two false alarm rates (relating to non-target or spoof trials). 
Priors and costs for a-DCF are listed below. 
\begin{equation}
\pi_\text{tar}=0.9405, \pi_\text{non}=0.0095, \pi_\text{spf}=0.05
\end{equation}
\begin{equation}
C_\text{miss}=1, C_\text{fa}=10,  C_\text{fa,spoof}=10
\end{equation}
Reference implementation of the metric is included in the evaluation package in GitHub repository mentioned in Section~\ref{sec:code}.

\subsection{Rules}
\begin{itemize}
    \item 
    Participant who register in the SASV track cannot participate in the TTS track.
    \item 
    Participant CAN use SSL model as a system module, e.g., feature extraction. 
    \item Each test sample must be scored independently of each other; techniques such as domain adaption using the evaluation segments, or any use of evaluation data for normalisation (across multiple or all test samples) purposes is NOT allowed. 
    \item 
    As required by the ASVspoof challenges~\cite{delgado2024asvspoof}, participants must not make public comparisons to the results or rankings of competing teams.  This rule applies also to the disclosure of a participant's own \emph{ranking} which is, by definition, a comparison to the results and rankings of competing teams. It applies also to claims of being the `challenge winner', `top-ranked system', `leading team', or any similar expressions which may be interpreted in any way as a comparison to the results of other participants.  Participants can, however, publish their \emph{own results}.  
    A summary of the challenge results will also be prepared by the organizers. Any participant wishing to retain anonymity should provide an anonymous team identifier with their registration. 
    \item 
    Participants commit to respecting the guidelines detailed earlier in this document.  Track, condition and data/resource use guidelines should also be interpreted as challenge rules.  If a participants is uncertain as to whether any particular data or resource is permitted, they are advised to seek clarification from the organizers.
\end{itemize}

The organizers reserve the right to exclude systems from ranking and to exclude teams from future participation in WildSpoof in the event that the above rules are not adhered to.

\section{Baseline and metrics implementations}
\label{sec:code}
The baseline system for the TTS track is Grad-TTS~\cite{popov2021gradtts} (as acoustic model) and DiffWave~\cite{kong2021diffwave} (as vocoder)\footnote{Code and pretrained checkpoints are available at \url{https://github.com/wildspoof/TTS_baselines}.}. 
Participants are encouraged to refer to the TITW paper for a detailed description of the system and its performance, as well as comparisons with other reference methods.
For the SASV track, we provide an end-to-end integrated system~\cite{mun2023towards} as a baseline\footnote{Code and pretrained checkpoints are available at \url{https://github.com/wildspoof/SASV_baselines}.} and the details can be found in the SpoofCeleb paper~\cite{jung2025spoofceleb}. 



\section{Ethics}
WildSpoof promotes ethical research and responsible practices to improve ASV security. We encourage participants to develop robust solutions for detecting spoofing and deepfakes while protecting the interests of all involved. Participants must adhere to local data protection laws and conduct their research responsibly, being mindful of potential misuse. We expect you to promptly and responsibly disclose any vulnerabilities or weaknesses found in ASV technology to help improve security and prevent malicious use. The organizers do not condone hacking, unauthorized access, or the malicious creation of spoofs and deepfakes. We strictly prohibit the misuse of any knowledge or tools developed through WildSpoof.

\section{Registration}
Challenge participants are required to register their interest through the following form:

\begin{center}
\href{https://docs.google.com/forms/d/e/1FAIpQLSf8RsuyfRvaNE2TI_uR9vCUgeZP_XMg80d_VTGCfozCVGRFkQ/viewform}{Registration form}
\end{center}

\section{Schedule}
\label{sec:schedule}

\noindent A tentative schedule is as follows (AOE):

\begin{itemize}
\item Data release: \hspace*{\fill}open-sourced done
\item Eval plan release: \hspace*{\fill}August 23, 2025
\item Challenge registration deadline: \hspace*{\fill}November 01, 2025
\item Submission deadline: \hspace*{\fill}December 01, 2025 (23:59 AoE)
\item Results announcement: \hspace*{\fill}December 05, 2025
\item 2-page papers due (by invitation only): \hspace*{\fill}December 07, 2025
\item Camera-ready 2-page papers due: \hspace*{\fill}January 18, 2026
\end{itemize}


\printbibliography

\end{document}